\documentstyle[12pt]{article}

\def\appendix{\par
    \setcounter{section}{0}
    \setcounter{subsection}{0}
    \renewcommand{\theequation}{\Alph{section}.\arabic{equation}}
    \renewcommand{\thesection}{Appendix \Alph{section}
		\setcounter{equation}{0}  } 
}

\catcode`\@=11
\def\@versim#1#2{\lower 2\p@\vbox{\baselineskip\z@skip\lineskip+1\p@
    \ialign{$\m@th#1\hfil##\hfil$\crcr#2\crcr\sim\crcr}}}
\def\gapp{\mathrel{\mathpalette\@versim>}}
\def\lapp{\mathrel{\mathpalette\@versim<}}
\catcode`\@=12

\let\be=\beta

\let\Ga=\Gamma
\let\de=\delta

\let\ep=\varepsilon

\let\La=\Lambda
\let\ph=\varphi

\let\th=\theta

\let\p=\partial
\let\<=\langle
\let\>=\rangle

\let\txt=\textstyle
\let\dsp=\displaystyle

\def\eqn#1{(\ref{#1})}  
\def\Eqn#1{Eq.~(\ref{#1})}  

\def\beq{\begin{equation}}
\def\eeq{\end{equation}}
\def\ba{\begin{array}}
\def\bea{\begin{eqnarray}}
\def\ea{\end{array}}
\def\eea{\end{eqnarray}}

\def\comment#1{ \hbox{[{\it Comment suppressed here.}\/]} }
\def\hide#1{}

\newdimen\pmboffset
\def\oldpmb#1{\setbox0=\hbox{#1}%
 \copy0\kern-\wd0
 \kern\pmboffset\raise 1.732\pmboffset\copy0\kern-\wd0
 \kern\pmboffset\box0}

\def\starttext{
\setlength{\baselineskip}{ 17pt} 
\pagenumbering{arabic}
}

\def\half{{\txt{1\over 2}}}
\def\Half{{1\over 2}}

\def\quar{{\txt{1\over 4}}}

\def\ie{i.e.~}
\def\Vb{v}    
\def\rhob{{\bar\rho}}

\def\vv{{\bf v}}
\def\ww{{\bf w}}
\def\WW{{\bf W}}

\newenvironment{equationwithlabel}[1]
  {
  \begin{equation}\label{#1}}{\end{equation}}

\newcommand{\beql}[1]{\begin{equationwithlabel}{#1}}
\newcommand{\eeql}{\end{equationwithlabel}}

\begin{document}

\title{ Critical Exponents without the Epsilon Expansion}

\author{
        Mark Alford\thanks{
          E-mail address: {\tt alford@hepth.cornell.edu}}\\[0.5ex]
	Laboratory of Nuclear Studies,\\
	Cornell University,\\
	Ithaca, NY 14853\\[2ex]
}
\newcommand{\preprintno}{
  \normalsize
 CLNS 94/1279\\[2ex] {\tt hep-ph/9403324}
}

\date{March 21, 1994\\[3ex] \preprintno}

\begin{titlepage}
\maketitle
\def\thepage{}		

\begin{abstract}

We argue that the sharp-cutoff Wilson renormalization
group provides a powerful tool for the analysis of second-order
and weakly first-order phase transitions.
In particular, in a computation no harder than
the calculation of the 1-loop effective potential, we show
that the Wilson RG yields the fixed point couplings and
critical exponents of 3-dimensional $O(N)$ scalar field theory,
with results close to those obtained in high-order $\ep$-%
expansion and large-$N$ calculations. We discuss the
prospects for an even more precise computation.

\end{abstract}

\end{titlepage}

\renewcommand{\thepage}{\arabic{page}}

\starttext

Weakly first-order and second-order phase transitions have been
important for some time in cosmology. Particular attention has been
focused on the electroweak phase transition (EWPT), in an effort to
determine whether it is first-order and thereby capable of generating
the observed asymmetry between matter and antimatter in the universe
\cite{CKN}.  However, progress has been complicated by infrared
problems that invalidate the usual loop expansion \cite{AE}.  In
statistical mechanics, the usual remedy is to
use the $\ep$-expansion \cite{WF,ZJ}, and this has been
applied to the EWPT \cite{AY}.  However, it has been argued that the
$\ep$-expansion may be unreliable for the EWPT \cite{JMR}, since there
may be a fixed point in three dimensions that is not visible in an
expansion around four dimensions.  In this letter we will show how a
second-order phase transition can be analyzed by the ``Wilson'' or
``exact'' renormalization group (RG)
\cite{WilsonRG,Pol} directly in three dimensions, without recourse to the
$\ep$-expansion.  It appears straightforward to extend this method to
first-order phase transitions \cite{AMR}, which would enable us to
determine the order and the dynamics of the EWPT and other
cosmological phase transitions.

There has recently been a resurgence of interest in the Wilson RG
(\cite{HHetc}--\cite{AMR}). An important
contribution was made by Tetradis and Wetterich
\cite{TW}, who used the smooth-cutoff Wilson RG to calculate
critical exponents of a second-order phase transition, and found
impressive agreement with the traditional methods mentioned above.
Unfortunately, the use of the smooth cutoff greatly complicates the
computation, and numerical methods are required to determine the
differential flow equations, which are then solved numerically to
obtain the exponents.  We will find, in contrast, that the
sharp-cutoff Wilson RG yields the fixed point couplings and critical
exponents in a simple and almost completely analytic form, requiring
no more effort than the calculation of the one-loop effective
potential. Even with a radically truncated action, the resultant values
are within about 10\% of those found by the traditional methods. In
the appendix we discuss the problems that arise with both smooth and
sharp cutoffs in a more precise calculation.

The Wilson RG follows the flow of the effective action (free energy)
as degrees of freedom are integrated out in successively shrinking
momentum shells.  This corresponds to calculating the effective action
with an infra-red (IR) cutoff $\La$ for the loop corrections, and
sending $\La\to 0$.  The effective action then flows with $\La$, its
couplings obeying differential equations that we will study below. For
high temperature systems, we are interested in the flow for the
3 dimensional theory, and if there is a second-order phase transition
we expect to find a fixed point that is IR-attractive in all
coupling directions except one. This corresponds to the fact that one
linear combination of the $UV$ couplings (\ie the temperature) must be
fine-tuned to a critical value if the theory is to flow into the fixed
point in the IR.  For temperatures close to the critical one, the
effective action stays very close to the critical point for a large
range of $\La$, and this causes the singular dependence of many
quantities on $T-T_c$.

We will consider the $O(N)$ scalar field theory in 3 dimensions.  It
will turn out that on the critical surface the mass-squared flows to
zero from below, so the true minimum is at a nonzero field.  By the $O(N)$
symmetry we can choose this to lie in the $\phi_1$ direction, so we
shift the field $\phi_i = \Phi \de_{i1} + \ph_i$, and the
effective action for the shifted field $\ph$ is
\beql{I:ham}
\ba{rcl}
{\cal F} &=& \dsp \half Z \p_\mu \ph_i \p^\mu \ph_i + V(\rho)\\[1ex]
\rho &\equiv& \half \phi^i\phi_i = \rho_0 + \Phi\ph_1 + \half \ph_i\ph_i
\ea
\eeql
This effective action is parameterized by the wavefunction
renormalization $Z$, the position $\rho_0=\half \Phi^2$ of the minimum
of the potential, and the derivatives of $V$ at $\rho_0$, $V_n = \p^n
V / \p\rho^n(\rho_0)$. By definition of $\rho_0$, $V_1=0$.

The procedure for obtaining the Wilson RG equations is simply to
calculate the effective action in the usual way, but with an IR cutoff
$\La$ in the propagators.  We then differentiate with respect to $\La$
to obtain the flow of $Z$, $\rho_0$, and the $V_n$ with $\La$
\cite{Pol,Wet,BDM,Morris}.

The flow of $Z$ is given by the anomalous dimension $\eta$, which is
discussed in the Appendix.  For the sake of computational simplicity,
we have assumed that $Z$ is a constant, so $\eta=0$. This is expected
to be a good approximation, since $\ep$- and
large-$N$ expansions indicate $\eta\lapp 0.04$.
The flow of the effective potential just corresponds to
differentiating with respect to an IR cutoff $\La$ in the momentum
integral of the usual 1-loop formula \cite{Sid}, so we easily obtain
the flow equation, first written down by Nicoll, Chang, and Stanley
(Ref.~\cite{WilsonRG}, 1976):
\beql{I:Vflow}
\La \left.{\p V(\rho) \over \p\La}\right|_\rho = -{K_3\over 2} \La^3
   \biggl[ \ln\Bigl(Z \La^2 + V'(\rho) + 2\rho V''(\rho)\Bigr)
    + (N-1)\ln\Bigl(Z \La^2 + V'(\rho) \Bigr) \biggr]~,
\eeql
where $K_3 = 1/(2\pi^2)$.  We emphasize that although this looks like
a 1-loop result there are no higher loop terms: the two-loop
contribution from a given shell goes as $(d\La)^2$ and does not affect
the $\La$-derivative. The approximation in \Eqn{I:Vflow} is that it
uses a truncated effective action, only taking account of the flow of
the momentum-independent couplings (the effective potential).
Including momentum-dependent vertices would clearly improve the
accuracy of the results.

Since we expect the {\em dimensionless} couplings to flow to a fixed
point, we now scale out the dimensions and $Z$-dependence, defining a
dimensionless field and couplings:
\beql{II:rescale}
\dsp \rhob = \rho {Z \over \La}, \quad  \Vb(\rhob) = \La^{-3} V
= v_0 + v_1(\rhob-\rhob_0) + \half v_2(\rhob-\rhob_0)^2 + \cdots
\eeql
Substituting these into \Eqn{I:Vflow} we find the flow of the
dimensionless couplings at constant $\rhob$.  Remembering that $\rhob$
is also flowing, our final result is
\beql{II:flow}
\La {d \Vb \over d\La} =
\La{d \rhob_0 \over d\La} {\p \Vb \over \p \rhob}
- 3 \Vb + \rhob {\p \Vb \over \p\rhob}
- \Half K_3 \biggl\{ \ln\Bigl(1 + \Vb'+2\rhob\Vb''\Bigr)
    + (N-1) \ln\Bigl( 1 + \Vb' \Bigr) \biggr\}.
\eeql
We determine the flow of $\rhob_0$ by noting that by definition of
$\rhob_0$, $v_1=0$ for all $\La$, so the flow equation for $v_1$
becomes a flow equation for $\rhob_0$:
\beql{rhoflow}
\La {d\rhob_0\over d\La} = -{1\over v_2} \La {\p v_1 \over \p \La}
= -\rhob_0 + \Half K_3 \biggl( (N-1)
+ {3 + 2 \rhob_0 v_3/v_2 \over 1 + 2\rhob_0 v_2} \biggr).
\eeql
We now expand \Eqn{II:flow} in powers of $(\rhob-\rhob_0)$ to find
the flow of $v_2$, $v_3$, etc.
\beql{vflow}
\ba{rcl}
\dsp \La {d v_2 \over d\La} &=&\dsp  v_3 \La{d \rhob_0 \over d\La}
 -  v_2 +\rhob_0 v_3 \\[2ex]
&-&\!\! \dsp \Half K_3 \biggl( (N-1)(v_3 - v_2^2 )
 - { (3 v_2 + 2 \rhob_0 v_3)^2 \over (1 + 2\rhob_0 v_2)^2} +
{5 v_3 + 2 \rhob_0 v_4 \over 1 + 2\rhob_0 v_2} \biggr), \\[2ex]
\dsp \La {d v_3 \over d\La} &=&\dsp v_4 \La{d \rhob_0 \over d\La}
 +\rhob_0 v_4  \\[2ex]
&-&\!\! \dsp \Half K_3 \biggl( (N-1)( v_2^3 - 3 v_2 v_3 + v_4)
+ { 2(3v_2 + 2 \rhob_0 v_3)^3 \over (1 + 2\rhob_0 v_2)^3}\\[3ex]
&&\dsp \qquad - ~~ 3 {(3v_2 + 2\rhob_0 v_3)(5v_3 + 2\rhob_0 v_4)
 \over (1 + 2\rhob_0 v_2)^2}
+ {7 v_4 + 2 \rhob_0 v_5 \over (1 + 2\rhob_0 v_2)}\biggr).
\ea
\eeql
Expanding to arbitrary order to find the flow of $v_n$ is a matter of
tedious but straightforward algebra.

To find the fixed point we set all flows to zero and solve
Eqs.~\eqn{rhoflow} and \eqn{vflow} for
the fixed point couplings $\rhob_0^*, v_2^*, v_3^* \ldots
v_n^*$. If our results are to be trustworthy then the fixed point
couplings must settle down at constant values as we relax the
truncation, \ie as $n\to\infty$. Fortunately, finding the fixed point
always involves solving {\em two} simultaneous equations, no matter
what the value of $n$, since the $\rhob_0$ equation eliminates $v_3^*$
in terms of $\rhob_0^*$ and $v_2^*$, the $v_2$ equation eliminates
$v_4^*$, and so on.  We finally set $v_{n+1}^*$ and $v_{n+2}^*$ to
zero, obtaining two simultaneous (polynomial) equations for
$\rhob_0^*$ and $v_2^*$.  Solving these is the only part of the
calculation which we have to perform numerically. The results for the
two scalar field case ($N=2$) are shown in Table~1 as a function of
$n$.  (They were obtained using the symbolic manipulation program {\em
Mathematica} to analytically expand the flow equations and
numerically find $\rhob_0^*$ and $v_2^*$.)  We see impressive
convergence of the couplings and the critical exponent $\be$
(calculated below) as the truncation of the potential is relaxed.

\begin{table}
\def\st{\rule[-1.5ex]{0em}{4ex}} 
\begin{center}
\begin{tabular}{|c|c|c|c|c|c|c|c|c|c|c|}\hline
\st $n$ & $\be$ & $\rhob_0^*$ & $v_2^*$ & $v_3^*$ & $v_4^*$ & $v_5^*$
& $v_6^*$ & $v_7^*$ & $v_8^*$ \\
\hline
\st 3 & 0.401 & 0.0716 & 9.70 & 92.2& 0   &  0    & 0     & 0      & 0 \\
\st 4 & 0.390 & 0.0692 & 10.9 & 106 & 303 &  0    & 0     & 0      & 0 \\
\st 5 & 0.385 & 0.0686 & 11.1 & 107 & 316 & -4660 & 0     & 0      & 0 \\
\st 6 & 0.384 & 0.0685 & 11.2 & 108 & 335 & -4710 & 58400 & 0      & 0 \\
\st 7 & 0.384 & 0.0684 & 11.2 & 108 & 335 & -4720 & 58400 & -64700 & 0 \\
\st 8 & 0.384 & 0.0685 & 11.2 & 108 & 333 & -4730 & 53700 & -98400 & -37M \\
\hline
\end{tabular}
\end{center}
\caption{The $N=2$ fixed point couplings and critical exponent $\be$,
calculated using $n$ evolution equations, \ie including vertices
up to $\phi^{2n}$. Note the striking convergence
with increasing $n$. ($M \equiv 10^6$). }
\end{table}

To calculate critical exponents, consider what happens
when $T$ is not quite equal to $T_c$, so that the dimensionless couplings
almost flow into the fixed point, but turn away at the last moment
and explode exponentially away from it.
{}From \Eqn{rhoflow} we see that when $\rhob_0$ becomes of order 1,
at some $\La=\La_f$,
the noncanonical scaling term (multiplied by $K_3$) becomes
negligible, and from then on $\rhob_0$ grows canonically,
so $\rho_0$ tends to a constant:
\beql{Laf}
\rho_0 \sim \La_f
\eeql
The critical exponent $\be$, defined by $\rho_0\sim (T-T_c)^{2\be}$,
can then be found by examining the flows in the neighborhood of the
fixed point. Write the couplings as a vector $\vv$, with components
$v_i-v^*_i$ (let $v_1$ be $\rhob_0$ for this purpose); then the flows
can be linearized:
\beq
\La {d \vv \over d \La} = \WW \cdot \vv
\eeq
so that
\beql{linflow}
W_{ij} = \left[{d \over d v_j} \La {d (v_i\!-\!v_i^*) \over d \La}
\right]_{v=v^*}
\eeql
We will find that $\WW$ has one positive eigenvalue $w_1$, corresponding to
the repulsive direction, with eigenvector $\ww_1$. It follows that
\beq
\vv(\La) = A(T-T_c) \La^{-w_1} \ww_1 + \cdots
\eeq
The couplings leave the vicinity of the fixed point when the coefficient
of $\ww_1$ becomes of order 1, so $A(T-T_c) \La_f^{-w_1} \sim 1$.
By \Eqn{Laf}, we otain the critical exponent $\be$:
\beql{critex}
\rho_0 \sim (T-T_c)^{1/w_1},\qquad
 \hbox{so}\quad \beta = 1/(2w_1),
\eeql
where $w_1$ is the largest eigenvalue of the linearized flow equations
\eqn{linflow} at the fixed point.  We can now calculate the fixed
point couplings, and the critical exponent $\be$ for any $N$. (Except
for the anomalous dimension, the other exponents follow by scaling
relations.)  The results are given in Table 2.

\begin{table}
\def\st{\rule[-1.5ex]{0em}{4ex}} 
\begin{center}
\begin{tabular}{|c|c|c|c|c|c|c|}\hline
\st N & $\beta$ & $\rhob_0^*$ & $v_2^*$ & $v_3^*$ & $v_4^*$  \\
\hline
\st   1 & 0.345 [0.327(2)] &  0.0471 &  13.5 &  177 &  955 \\
\st   2 & 0.385 [0.349(4)] & 0.0686 & 11.1 & 107 & 316 \\
\st   3 & 0.413 [0.368(4)] & 0.0917 & 9.14 & 66.7 & 134 \\
\st  10 & 0.473 [0.449] & 0.266  & 3.60 & 9.06 & 6.44 \\
\st  20 & 0.487 [0.477] & 0.518 & 1.90 & 2.43 & 0.910 \\
\st  50 & 0.495 [0.491] & 1.28 & 0.776 & 0.405 & 0.0626 \\
\st 100 & 0.497 [0.496] & 2.54 & 0.391 & 0.103 & 0.00801 \\
\hline
\end{tabular}
\end{center}
\caption{The $O(N)$ fixed point couplings and critical exponents for
various $N$. The values in square brackets are the current
best estimates, based on $\ep$-expansion to order $\ep^5$
($N \leq 3$) \protect\cite{ZJ},
or large $N$ to order $1/N^2$ ($N \geq 10$) \protect\cite{KT}.
}
\end{table}


We see that there is striking agreement between the sharp-cutoff
Wilson RG, and conventional $\ep$-expansion and large-$N$ results,
particularly so considering the crudity of our approximation, in which
all momentum dependence in the effective action has been ignored.  These
results also agree closely with those obtained numerically by Tetradis
and Wetterich \cite{TW} via the smooth-cutoff Wilson RG.  It seems
reasonable to expect that the inclusion of momentum-dependent vertices
would enhance the accuracy of the results still further. However, as
we show in the appendix, this is not a completely straightforward
procedure. In particular, the smooth-cutoff calculation \cite{TW} of the
anomalous dimension appears to depend on the profile of the cutoff,
rendering its physical significance unclear.

For cosmological applications, it is necessary
to formulate a gauge-invariant version of the momentum cutoff, in
order to study gauge theories with weakly first-order phase
transitions such as the electroweak phase transition.
Progress in this direction has already been reported \cite{gauge}.
It would also be interesting to see if the sharp-cutoff Wilson RG could be
applied with equal success to second order phase transitions in other
statistical mechanical models, or to calculation of the effective
potential in other field theories where perturbation theory breaks
down in the IR, such as multiple scalar field theories with radiatively
induced first-order phase transitions of the Coleman-Weinberg type
\cite{AMR}.

\vspace{36pt}
\centerline{\bf Acknowledgments}
I would like to thank John March-Russell and Tim Klassen for
discussions.

\appendix
\section{The anomalous dimension}

The anomalous dimension is given by the flow of the wavefunction
renormalization $Z$, so it is convenient to express the effective
action in terms of $\ph_i$ rather than $\rho$:
\beql{A:pot}
V(\rho) = \Half V_2\Bigl(
\Phi^2 \ph_1^2 + \Phi\ph_1 \ph_i\ph_i + \quar\ph_i\ph_i \ph_j\ph_j \Bigr)
+ \cdots
\eeql
The first term is the mass of the radial mode, which comes into
our flow formulae in its dimensionless form, as $2\rhob_0 v_2$.
The next term provides a cubic interaction between the Goldstone
and radial modes. This is the term that, via the graph shown in
Fig.~1, will drive the cutoff-dependence of the wavefunction
renormalization $Z$ of the Goldstone modes.

\begin{figure}

\protect\[ 
\Ga^{(2)}_{2,2}(k) = \quad
  \begin{picture}(100,45)(0,7)
    \thicklines
    \put(10,10){\line(1,0){20}}
    \put(50,10){\circle{40}}
    \put(70,10){\line(1,0){20}}   

    \put(20,14){\makebox(0,0){$\rightarrow$}}  
    \put(80,14){\makebox(0,0){$\rightarrow$}}

    \put(20,25){\makebox(0,0){$k$}}   
    \put(80,25){\makebox(0,0){$k$}}
    \put(4,10){\makebox(0,0){$2$}}   
    \put(97,10){\makebox(0,0){$2$}}

    \put(50,22){\makebox(0,0){1}}   
    \put(50,-2){\makebox(0,0){2}}

    \put(50,35){\makebox(0,0){$\leftarrow$}} 
    \put(50,-15){\makebox(0,0){$\rightarrow$}}

    \put(50,45){\makebox(0,0){$q$}}  
    \put(50,-25){\makebox(0,0){$q+k$}}
  \end{picture}
\protect\] 
\vspace{1ex}
\caption{
The diagram that contributes to the coefficient of
$k^2$ in the 2-point 1PI Green's function, \ie to the
running of $Z$.
}
\end{figure}

We calculate $Z=\half\p^2\Ga^{(2)}_{2,2}/\p k^2$ by evaluating this
diagram in the usual way, but with an IR cutoff $\th_\ep(p^2 - \La^2)$
(which becomes a step function as $\ep\to 0$) with every propagator.
Differentiating with respect to $\La$,
\beql{A:diag}
\ba{r @{} c @{} l}
\dsp \eta_g = {\La\over Z} {\p Z \over \p\La}
= (\Phi V_2)^2 {\La \over Z} \Half {\p^2 \over \p k^2}
&\dsp\int & \dsp{d^3q\over (2\pi)^3} \left\{
{ \de_\ep(q-\La) \over Z q^2 + 2 \rho_0 V_2 }
  {\th_\ep\bigl( (q+k)^2-\La^2) \over Z(q+k)^2 } \right. \\[3ex]
&+&\dsp\left. { \de_\ep(q-\La) \over Z q^2 }
  {\th_\ep\bigl( (q+k)^2-\La^2) \over Z(q+k)^2 +2 \rho_0 V_2 } \right\}
\ea
\eeql
This formula is to be evaluated in the limit $k\to 0$, and the sharp
cutoff limit $\ep\to 0$; but in which order?  Clearly if we send $k\to
0$ first then the $k$-derivatives convert the $\th_\ep$-functions into
$\de_\ep$-functions (and their derivatives) centered at $q=\La$, and
we obtain a result that is sensitively dependent on $\ep$, the profile
of the cutoff, and diverges as $\ep\to 0$. Moreover, the expansion in
$k$ is only valid for $k\lapp \ep$ \cite{Morris}.  This is not a
problem with the sharp cutoff limit, as has been suggested
\cite{UE,Wet}, since it is just as bad for the smooth cutoff to give
profile-dependent results as it is for the sharp cutoff to give
infinite ones. Rather, it implies that the order of limits is wrong.
Physically, we want our effective action to be valid for momenta spanning as
much of the range $0$ to $\La$ as possible. This is achieved by
sending $\ep\to 0$ first, and we then obtain $\ep$-independent results
for any $k\gg \ep$, just as one would expect.

A more serious problem is that the resultant 2-point function contains
a non-local $|k|$ term, as well as the expected $k^2$ term (see also
Ref.~\cite{Wet}).  This occurs because we have the same infrared
cutoff on two different loop momenta, and only part of the loop is
integrated out.  This does not mean that the Greens functions, which
are physically measurable, are non-local.  The effective action with
an IR cutoff is not a physically measurable quantity, and so it may
indeed contain non-local terms \cite{Morris}.  It is not clear whether
we should include a $|k|$ term in our effective action, with a
separate flow equation for its coefficient, or whether we can amend
the definition of the coarse-grained effective action in such a way as
to eliminate non-local terms. This issue clearly deserves further
study.


\end{document}